\documentclass[twocolumn,%superscriptaddress,
amsmath,amssymb,aps,pra]{revtex4-1}

\usepackage{graphicx}
\usepackage{dcolumn}
\usepackage{bm}

\usepackage[colorlinks,urlcolor=blue,citecolor=blue,linkcolor=blue]{hyperref}

\usepackage{comment}
\usepackage{color}
\usepackage{physics}

\newcommand{\up}{\uparrow}
\newcommand{\down}{\downarrow}
\renewcommand{\k}{{\bf k}}
\newcommand{\p}{{\bf p}}

\newcommand{\q}{{\bf q}}
\newcommand{\Q}{{\bf Q}}

\newcommand{\0}{{\bf 0}}

\newcommand{\ef}{E_F}
\newcommand{\kf}{k_F}
\newcommand{\eb}{\varepsilon_{\rm B}}
\newcommand{\eq}{\epsilon_{\q}}
\newcommand{\ek}{\epsilon_{\k}}
\newcommand{\ep}{\epsilon_{\p}}

\newcommand{\nn}{\nonumber}
\newcommand{\beq}{\begin{equation}}
\newcommand{\eeq}{\end{equation}}

\begin{document}

\title{Thermodynamic signatures of the polaron-molecule transition in a Fermi gas}

\author{Meera M. Parish}
\affiliation{School of Physics and Astronomy, Monash University, Victoria 3800, Australia}
\affiliation{ARC Centre of Excellence in Future Low-Energy Electronics Technologies, Monash University, Victoria 3800, Australia}

\author{Haydn S. Adlong}
\affiliation{School of Physics and Astronomy, Monash University, Victoria 3800, Australia}
\affiliation{ARC Centre of Excellence in Future Low-Energy Electronics Technologies, Monash University, Victoria 3800, Australia}

\author{Weizhe Edward Liu}
\affiliation{School of Physics and Astronomy, Monash University, Victoria 3800, Australia}
\affiliation{ARC Centre of Excellence in Future Low-Energy Electronics Technologies, Monash University, Victoria 3800, Australia}

\author{Jesper Levinsen}
\affiliation{School of Physics and Astronomy, Monash University, Victoria 3800, Australia}
\affiliation{ARC Centre of Excellence in Future Low-Energy Electronics Technologies, Monash University, Victoria 3800, Australia}

\date{\today}

\begin{abstract}
We consider the highly spin-imbalanced limit of a two-component Fermi gas, where there is a small density of $\down$ impurities attractively interacting with a sea of $\up$ fermions. In the single-impurity limit at zero temperature, there exists the so-called polaron-molecule transition, where the impurity sharply changes its character by binding a $\up$ fermion at sufficiently strong attraction. Using a recently developed variational approach, we calculate the thermodynamic properties of the impurity, and we show that the transition becomes a smooth crossover at finite temperature due to the thermal occupation of excited states in the impurity spectral function. However, remnants of the single-impurity transition are apparent in the momentum-resolved spectral function, which can in principle be probed with Raman spectroscopy. We furthermore show that the Tan contact exhibits a characteristic non-monotonic dependence on temperature that provides a signature of the zero-temperature polaron-molecule transition. 
For a finite impurity density, we argue that descriptions purely based on the behavior of the Fermi polaron are invalid near the polaron-molecule transition, since correlations between impurities cannot be ignored. In particular, we show that the spin-imbalanced system undergoes phase separation at low temperatures due to the strong attraction between $\up\down$ molecules induced by the Fermi sea. Thus, we find that the impurity spectrum and the induced impurity-impurity interactions are key to understanding the phase diagram of the spin-imbalanced Fermi gas. 
\end{abstract}

\maketitle

\section{Introduction}

The problem of a mobile impurity immersed in a Fermi gas is important for a variety of systems ranging from neutron stars~\cite{Kutschera1993} to trapped ultracold atoms~\cite{Massignan2014} to the absorption spectra in doped semiconductors~\cite{Sidler2017,Efimkin2017,Efimkin2018}.
In this scenario, the impurity attractively interacts with the surrounding fermions and becomes dressed by excitations of the Fermi gas to form a quasiparticle --- also termed a Fermi polaron --- with modified properties such as a larger effective mass~\cite{Massignan2014}. 
The Fermi polaron has an exceptionally clean realization in ultracold atomic gases~\cite{Schirotzek2009,Nascimbene2010,Kohstall2012,Koschorreck2012,Zhang2012,Cetina2015,Cetina2016,Scazza2017,Yan2019,Oppong2019,Ness2020}, where the impurity-fermion interactions can be precisely tuned and the temperature can be varied from the quantum degenerate to the classical regime. Moreover, the case of fermionic impurities corresponds to the limit of extreme population imbalance in a two-component Fermi gas~\cite{Chevy2006,Radzihovsky2010,Chevy2010}, where the impurities form the minority ($\down$) component and the Fermi medium consists of the majority ($\up$) fermions. Such a spin-imbalanced Fermi gas provides a model system for exploring fermion pairing phenomena and exotic superfluid phases~\cite{Casalbuoni2004,Kinnunen2018}.

Of particular interest is the so-called polaron-molecule transition at zero temperature~\cite{Prokofev2008,Mora2009,Punk2009,Combescot2009,Bruun2010,Vlietinck2013}, where the $\down$ impurity suddenly binds a $\up$ fermion at sufficiently strong attraction and forms a dressed $\up\down$ molecule or dimer. Most notably, the dressed dimer has a vanishing overlap with the bare non-interacting impurity and thus corresponds to a radically different quasiparticle from the original dressed impurity or polaron. 
However, there has been much debate about the fate of this single-impurity transition once there is a finite impurity density or a finite temperature~\cite{Chevy2010,Frank2018,Tajima2018,Ness2020,Cui2020}. Most recently, it has been proposed that polarons and molecules coexist in the highly spin-imbalanced Fermi gas  
at low temperatures~\cite{Ness2020,Cui2020}, 
which has implications for the phase diagram of this system. 

In this paper, we reveal how the behavior of the polaron-molecule transition at finite temperature is intimately linked to the structure of the impurity spectral function. 
In particular, at zero temperature, the polaron-molecule transition is connected to the appearance of two degenerate minima in the momentum-resolved spectrum, as first elucidated in Ref.~\cite{Mathy2011}.  
Using a recently developed variational approach~\cite{Liu2019}, we show how the sharp single-impurity transition becomes a smooth crossover at finite temperature due to the thermal occupation of excited states in the impurity spectral function and the smearing of the $\up$ Fermi surface. 
However, signatures of the zero-temperature polaron-molecule transition can still be observed in thermodynamic quantities such as the Tan contact~\cite{Tan2008}, which we find monotonically decreases with temperature on the molecule side of the transition, while displaying a characteristic non-monotonic temperature dependence on the polaron side, consistent with recent experimental measurements for unitarity-limited impurity-fermion interactions~\cite{Yan2019}.
We also discuss how the structure of the spectral function can in principle be probed with Raman 
spectroscopy, and that the polaron-molecule transition can be particularly clearly resolved if impurities are ``injected'' into the interacting state, rather than being ``ejected'' from the interacting system as in a recent experiment~\cite{Ness2020}.

While thermal fluctuations destroy the single-impurity transition, we argue that a finite impurity density typically results in a first-order phase transition between superfluid and Fermi liquid phases at sufficiently low temperatures. 
In particular, we show that there is a strong attraction between dressed $\up\down$-dimers  induced by the surrounding Fermi gas, such that the dimer superfluid is unstable towards phase separation at zero temperature. 
Therefore, induced interactions between dressed impurities are crucial for describing the low-temperature phase diagram of the spin-imbalanced Fermi gas.

This paper is organized as follows. In Sec.~\ref{sec:model}, we introduce the Hamiltonian for the spin-imbalanced Fermi gas and we discuss the fundamental properties of different spectroscopic probes in the single-impurity limit. 
We furthermore outline the variational method that we use to calculate the impurity spectral function. In Sec.~\ref{sec:spectral}, we discuss signatures of the polaron-molecule transition in the impurity spectral function as well as the impurity free energy and the Tan contact. 
We also show how Raman injection spectroscopy can enhance the signal of dressed molecule states in the spectrum. 
In Sec.~\ref{sec:phase}, we discuss the behavior of the Fermi polaron in relation to the spin-imbalanced phase diagram, and we show how phase separation arises from the induced interactions between dressed molecules in the Fermi gas. We conclude in Sec.~\ref{sec:conc}.
The theoretical details for Raman spectroscopy of polarons is contained in Appendix~\ref{app:Raman}. 

\section{Model} \label{sec:model}
We model the 
two-component 
Fermi gas with short-range interactions using the Hamiltonian
\begin{align}
    \hat{H} = \sum_{\k\sigma} \ek \hat{c}^\dag_{\k\sigma} \hat{c}_{\k\sigma} + g \sum_{\k\k'\q} \hat c^\dag_{\k\up} \hat c^\dag_{\k' \down}\hat c_{\k'-\q \down} \hat c_{\k+\q \up} ,
    \label{eq:Hamiltonian}
\end{align}
where we work in units where the system volume and the Planck and Boltzmann constants are unity. The operator $\hat c^\dag_{\k\sigma}$ creates a fermion with momentum $\k$ and spin index $\sigma$, where the values $\sigma=\up,\down$ indicate distinct hyperfine states. The fermions have mass $m$ and dispersion $\ek = |\k|^2/2m\equiv k^2/2m$. The interactions only occur between distinguishable particles, and they are taken to be of zero range and characterized by a strength $g$ as in the case of a broad Feshbach resonance. We can relate the interaction strength $g$ to the scattering length $a$ via
\begin{align}
    \frac1g=\frac m{4\pi a}-\sum_\k^\Lambda\frac1{2\ek},
\end{align}
where $\Lambda$ is an ultraviolet cutoff on the relative momentum of the scattering particles.

\subsection{Probes of polaron physics} \label{sec:probes}

In the following, we consider the scenario of a majority spin-$\up$ Fermi sea and a small minority component of spin-$\down$ impurities. The properties of the dressed impurities (or polarons) are encoded in the spectral function~\cite{FetterBook} 
\begin{align} \label{eq:spec}
    A(\p,\omega)=-\frac1\pi\mathrm{Im}\left[G_\down(\p,\omega+i0)\right],
\end{align}
where $G_\down$ is the impurity Green's function at momentum $\p$ and energy $\omega$. We have included an imaginary infinitesimal  $+i0$  that shifts the  poles into the lower half of the complex plane since we are dealing with the retarded Green's function. The calculation simplifies in the limit of a single impurity atom, which is an accurate description of the system when the impurities are uncorrelated with each other. This is the case when the impurity density $n_\down$ is sufficiently low and/or the temperature $T$ is sufficiently high, as we discuss in Section~\ref{sec:phase}. Note that the spectral relationships investigated in this section are independent of dimensionality. 

We now outline the various spectroscopic protocols that can probe the polaron. The most common is radio-frequency (rf) spectroscopy, which comes in two flavors. In \textit{injection} spectroscopy, impurities initially occupy an auxiliary state, and (in the ideal case) this state is non-interacting with the medium. Within linear response theory, the rate at which impurities with definite momentum $\p$ are transferred from this auxiliary state to the spin-$\down$ state is proportional to the injection spectral function~\cite{Haussmann2009}
\begin{align} \label{eq:Ainj}
    A_{\rm inj}^{\rm rf}(\p,\omega)&=A(\p,\omega+\ep),
\end{align}
where the frequency $\omega$ is measured relative to the bare transition. Most cold-atom experiments are not momentum resolved, and instead probe the total spectral function
\begin{align} \label{eq:Iinj}
    I_{\rm inj}^{\rm rf}(\omega) & = \sum_\p n_{\rm B}(\p) A_{\rm inj}^{\rm rf}(\p,\omega),
\end{align}
which is averaged over the initial impurity momenta. We take the impurities to be uncorrelated, and therefore they are initially described by a Boltzmann distribution $n_{\rm B}(\p)=e^{-\beta \ep}/Z_{\rm imp}$, where $Z_{\rm imp}= \sum_\p e^{-\beta\ep}$ is the single-impurity partition function, and $\beta=1/T$ is the inverse temperature. Note that Eqs.~\eqref{eq:Ainj} and~\eqref{eq:Iinj} neglect any prefactors related to the strength of the rf or optical field.

An alternative protocol is ejection spectroscopy, where the impurity is ejected from the interacting spin-$\down$ state to the auxiliary state. The transfer rate into impurity states of momentum $\p$ is now proportional to the ejection spectral function $A_{\rm ej}^{\rm rf}(\p,\omega)$. It turns out that, in the single-impurity limit, this is related to the injection spectral function via the fundamental relationship~\cite{Liu2020a,Liu2020b}
\begin{align}
\label{eq:ejrf}
    A_{\rm ej}^{\rm rf}(\p,\omega)=e^{\beta \,\Delta F}e^{\beta\omega}n_{\rm B}(\p)A_{\rm inj}^{\rm rf}(\p,-\omega).
\end{align}
The proportionality factor depends on the impurity free energy $\Delta F$, defined as the difference between the free energy of the interacting and non-interacting systems, $\Delta F=F-F_0$. 
Similarly, the total ejection rate can be related to the total (momentum-averaged) injection spectral function~\cite{Liu2020a,Liu2020b}:
\begin{align}
    I_{\rm ej}^{\rm rf}(\omega)=\sum_\p A^{\rm rf}_{\rm ej}(\p,\omega)=e^{\beta\,\Delta F}e^{\beta \omega}I_{\rm inj}^{\rm rf}(-\omega).
\end{align}

Recently, the Fermi polaron has also been probed via Raman spectroscopy~\cite{Ness2020}. Like rf spectroscopy, this couples the $\down$ impurity state to an auxiliary state. However, unlike rf spectroscopy, Raman spectroscopy is effectively a two-photon process, and thus, apart from imparting a frequency shift $\omega$, it also imparts a momentum $\q$ to the impurity. Within linear response, the injection transfer rate for impurities initially with momentum $\p-\q$ in the auxiliary state is (see Appendix~\ref{app:Raman})
\begin{align} \label{eq:spec-rel}
    A_{\rm inj}^{\rm R}(\p;\q,\omega) & = A(\p,\omega+\epsilon_{\p-\q}),
\end{align}
where the final $\down$ dressed impurity has momentum $\p$. The momentum averaged transfer rate is then (see also Ref.~\cite{Haussmann2009})
\begin{align} \label{eq:Iinj-q}
    I_{\rm inj}^{\rm R}(\q,\omega) & = \sum_\p n_{\rm B}(\p-\q)A_{\rm inj}^{\rm R}(\p;\q,\omega).
\end{align}
In the limit $\q\to0$, these expressions reduce to those for rf spectroscopy. 

Raman spectroscopy may also be used to eject particles from the interacting impurity state, as was done in the recent experiment~\cite{Ness2020}. Performing a calculation similar to that in Refs.~\cite{Liu2020a,Liu2020b}, we obtain the relationship between injection and ejection Raman spectral functions: 
\begin{align} \label{eq:det-bal1}
    A_{\rm ej}^{\rm R}(\p;\q,\omega) = e^{\beta \,\Delta F}e^{\beta\omega}n_{\rm B}(\p-\q)A_{\rm inj}^{\rm R}(\p;\q,-\omega),
\end{align}
where the final impurity state has momentum $\p-\q$. Thus, the total ejection rate in Raman spectroscopy is
\begin{align}\label{eq:det-bal2}
    I_{\rm ej}^{\rm R}(\q,\omega) = \sum_\p A_{\rm ej}^{\rm R}(\p;\q,\omega) = e^{\beta \,\Delta F}e^{\beta\omega}I_{\rm inj}^{\rm R}(\q,-\omega).
\end{align}
For a derivation of these relationships, see Appendix~\ref{app:Raman}.

The Raman spectral functions satisfy the usual sum rules (see, e.g., Ref.~\cite{Schneider2010}) independently of $\q$:
\begin{subequations}
\begin{align} \label{eq:spec-sum1}
    \int d\omega \,A_{\rm inj}^{\rm R}(\p;\q,\omega) & = \int d\omega \,A(\p,\omega) =1,\\ \label{eq:spec-sum2}
    \int d\omega \,A_{\rm ej}^{\rm R}(\p;\q,\omega) & =n_{\rm int}(\p),
\end{align}
\end{subequations}
where $n_{\rm int}(\p)$ is the impurity momentum distribution in the interacting state (for details, see Appendix~\ref{app:Raman}). We also have
\begin{subequations}
\begin{align} \label{eq:sum1}
    \int d\omega \,I_{\rm inj}^{\rm R}(\q,\omega)
    & = \int d\omega \,I_{\rm inj}^{\rm rf}(\omega)= 1, \\
    \int d\omega \,I_{\rm ej}^{\rm R}(\q,\omega)
    & = \int d\omega \,I_{\rm ej}^{\rm rf}(\omega)=1.\label{eq:sum2}
\end{align}
\end{subequations}
Note that Eq.~\eqref{eq:sum2} is only valid in systems where the internal state of the impurity cannot be changed by the impurity-medium interactions~\cite{Liu2020b}. This condition is satisfied in our model~\eqref{eq:Hamiltonian}.

\subsection{Variational approach}
To approximate the impurity spectral function, we start by considering the impurity Green's function in the time domain:
\begin{align} \nn
    {\cal G}_\down(\p,t) = &
    \int d\omega\, e^{-i\omega t} G_\down(\p,\omega +i0) \\
    = & 
    -i \Theta(t)  \Tr\left[\hat{\rho}_0 \, 
    \hat{c}_{\p \down}(t) \hat{c}^\dag_{\p \down}(0) \right] ,
\end{align}
Here, $\Theta(t)$ is the Heaviside function, and we have introduced the time-dependent impurity 
operator $\hat{c}_{\p \down}(t) = e^{i\hat{H}t} \hat{c}_{\p \down} e^{-i\hat{H}t}$. The trace is taken over medium-only states, with the density matrix for the medium 
\begin{align}
    \hat{\rho}_0 = \frac{e^{-\beta \hat{H}_{\rm med}}}{\Tr\left[e^{-\beta \hat{H}_{\rm med}}\right]}.
\end{align}
$\hat{H}_{\rm med} = \sum_\k (\ek -\mu) \hat{c}^\dag_{\k \up} \hat{c}_{\k \up}$ is the medium-only Hamiltonian, and $\mu$ is the chemical potential of the $\up$ Fermi gas.

Following Ref.~\cite{Liu2019}, we consider an approximate impurity operator with at most one particle-hole excitation,
\begin{align} \label{eq:polT}
    \hat{c}_{\p \down}(t) & \simeq  \alpha_{\p;0}(t) \hat{c}_{\p \down}    
      + \sum_{\k \q} \alpha_{\p;\k\q}(t) \hat{c}_{\q\up}^\dagger \hat{c}_{\k\up} \,
    \hat{c}_{\p-\k+\q,\down} ,
\end{align}
where the variational parameters $\alpha_j(t)$ are complex functions of time.
Note that, unlike Ref.~\cite{Ness2020}, we only require a \textit{single} variational ansatz rather than two to capture the polaron-molecule transition and its evolution with temperature.
We then minimize the error 
\begin{align}
    \Delta_\p(t) = \Tr\left[\hat{\rho}_0 \hat{\varepsilon}_\p(t) \hat{\varepsilon}_\p^\dag(t)  \right],
\end{align}
with respect to each of the $\alpha_j(t)$, where the error operator $\hat{\varepsilon}_\p(t) \equiv i\partial_t \hat{c}_{\p\down}(t) - [\hat{c}_{\p\down}(t),\hat{H}]$ corresponds to the error incurred in the Heisenberg equation of motion.
Taking the stationary condition $\alpha_{\p;0}(t) = \alpha_{\p;0} e^{-iEt}$ and  $\alpha_{\p;\k\q}(t)  = \alpha_{\p;\k\q} e^{-iEt}$  yields the coupled equations~\cite{Liu2019}:
\begin{subequations} \label{eq:stationary}
\begin{align} 
    E \alpha_{\p;0} = & \, \ep \alpha_{\p;0} + g\sum_{\k\q} \alpha_{\p;\k\q} \, n_{\rm F}(\q) \left[1-n_{\rm F}(\k) \right] \\ \nn
    E \alpha_{\p;\k\q} = & \, \left(\epsilon_{\p +\q -\k} + \ek -\eq \right)\alpha_{\p;\k\q} + 
    g \alpha_{\p;0} \\
    & + g \sum_{\k'} \alpha_{\p;\k'\q} \left[1-n_{\rm F}(\k') \right] ,
\end{align}
\end{subequations}
with Fermi-Dirac distribution
\begin{align}
    n_{\rm F} (\k) = \Tr \left[\hat{\rho}_0 \, 
    \hat{c}^\dag_{\k \up} \hat{c}_{\k \up} \right] = \frac{1}{e^{\beta(\ek-\mu)}+1} .
\end{align}
Equation~\eqref{eq:stationary} corresponds to a matrix eigenvalue problem which can be solved to give eigenvectors $\{\alpha_{\p;0}^{(l)},\alpha_{\p;\k\q}^{(l)}\}$ and associated eigenvalues $E_\p^{(l)}$.
This finally yields the approximate impurity Green's function in the time domain
\begin{align}
    {\cal G}_\down(\p,t) = -i \Theta(t) \sum_l \big| \alpha_{\p;0}^{(l)}\big|^2 e^{-i E_\p^{(l)} t}.
\end{align}
By a Fourier transform, we arrive at the impurity Green's function in the frequency domain
\begin{align}
    G_\down(\p,\omega+i0) = \sum_l \frac{\big| \alpha_{\p;0}^{(l)}\big|^2}{\omega-E_\p^{(l)}+i0}, 
\end{align}
as well as the corresponding spectral function
\begin{align} \label{eq:spec2}
    A(\p,\omega) =  \sum_l \big|\alpha_{\p;0}^{(l)}\big|^2 \delta(\omega - E_\p^{(l)}) .
\end{align}

The spectral function of the Fermi polaron has been investigated with a variety of theoretical tools, including variational approaches at zero~\cite{Parish2016} and finite temperature~\cite{Liu2019}, $T$-matrix  
approximations~\cite{Punk2007,Massignan2008,Hu2018,Tajima2019,Mulkerin2019}, the functional renormalization group~\cite{Schmidt2011}, large-$N$ expansion~\cite{Veillette2008,Schneider2010},
and diagrammatic quantum Monte Carlo (QMC)~\cite{Goulko2016}. The variational approach where we consider a single particle-hole excitation of the Fermi sea is equivalent to a finite-temperature Green's function approach that only includes ladder diagrams~\cite{Liu2019}.

\section{Impurity spectral function and thermodynamics} \label{sec:spectral}

In this section, we focus on the limit of a single $\down$ impurity in a $\up$ Fermi sea. 
As discussed in Sec.~\ref{sec:probes}, this allows us to make precise statements about the impurity spectral function as well as the impurity thermodynamics via the free energy $\Delta F$. We take a fixed $\up$-fermion density $n_\up = \sum_\k n_{\rm F}(\k)$, with corresponding 
Fermi momentum $k_F = (6\pi^2 n_\up)^{1/3}$, Fermi energy $\ef =\kf^2/2m$, and Fermi temperature $T_F=\ef$. 
The relevant dimensionless parameters in the single-impurity limit 
are then the interaction strength $1/\kf a$ and the temperature $T/T_F$.

\subsection{Polaron-molecule transition}
In the zero-temperature limit, it has been established that the polaron undergoes a transition at sufficiently strong attraction $(1/k_F a)_0$, where the impurity binds a fermion to form a dressed dimer or molecule~\cite{Prokofev2008,Mora2009,Punk2009,Combescot2009,Bruun2010,Vlietinck2013}. To gain insight into this transition, it is instructive to consider the structure of the possible impurity states. Within the one particle-hole approximation, the polaron wave function at momentum $\p$ has the form~\cite{Chevy2006}
\begin{align} \label{eq:pol-wfn}
    \ket{\psi_\p} = \Big[\alpha_{\p;0}\hat{c}^\dag_{\p \down}  + \sum_{\k \q} \alpha_{\p;\k\q} \, \hat{c}^\dag_{\p-\k+\q,\down}  \hat{c}^\dag_{\k\up} \hat{c}_{\q\up} \Big] \ket{\rm FS} ,
\end{align}
where $\ket{\rm FS}$ is the Fermi sea of $\up$ fermions. Here we have $q\leq \kf < k$, and we can take the coefficients $\alpha$ to be real without loss of generality in the time-independent case. A key feature of the polaron state is that it has a finite residue $\alpha_{\p;0}^2$, which corresponds to the squared overlap with the non-interacting state $\hat{c}^\dag_{\p \down} \ket{\rm FS}$. Thus, the polaron appears as a delta function peak in the impurity spectral function at $T=0$ --- see Eq.~\eqref{eq:spec2}. 

On the other hand, the lowest-order form of the dimer at momentum $\Q$ is the Pauli-blocked state:
\begin{align} \label{eq:mol-wfn}
    \ket{\psi^M_\Q}
    = \sum_\k 
    \phi_{\Q;\k} \, \hat{c}^\dag_{\Q-\k,\down}  \hat{c}^\dag_{\k\up} 
    \ket{\rm FS}^\prime ,
\end{align}
where $\ket{\rm FS}^\prime$ corresponds to the Fermi sea with one less $\up$ fermion in order to conserve particle number among the different impurity states.  
It has previously been shown that the zero-momentum molecule $\ket{\psi^M_\0}$ becomes lower in energy than the zero-momentum polaron $\ket{\psi_\0}$ when $1/\kf a \geq (1/\kf a)_0\simeq 1.27$~\cite{Mora2009,Punk2009,Combescot2009}, whereupon the residue abruptly vanishes and the impurity quasiparticle radically changes its character. Including more particle-hole excitations of the Fermi sea only shifts this polaron-molecule transition to lower attraction, $(1/\kf a)_0 \simeq 0.88$~\cite{Prokofev2008,Mora2009,Punk2009,Combescot2009,Bruun2010,Vlietinck2013}; it does not destroy it. 

\begin{figure}
    \centering
    \includegraphics[width=\linewidth]{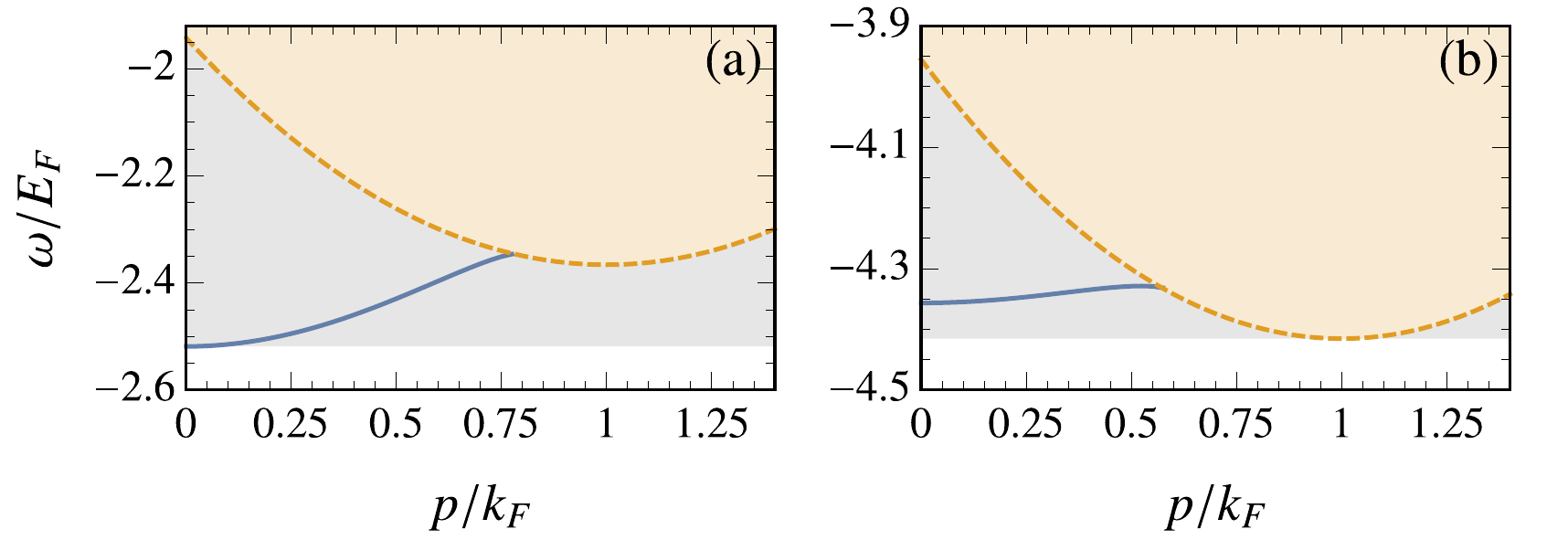}
    \caption{
    Polaron energy (solid blue) and the onset of the molecule-hole continuum (dashed orange) 
    for $1/k_F a = 1$ (a) and $1/k_F a = 1.4$ (b), corresponding to before and after the polaron-molecule transition, respectively.
    The energies are calculated using the zero-temperature ansatz in Eq.~\eqref{eq:pol-wfn},
    where the polaron-molecule transition occurs at 
    $(1/k_F a)_0 \simeq 1.27$ at this level of the approximation. 
    The exact spectrum has a continuum of states (gray shading) starting at the ground-state energy and spanning all momenta but with less spectral weight than the molecule-hole continuum (orange shading).}
    \label{fig:PolaronAndMoleculeHoleEdge}
\end{figure}

If we take $\ket{\rm FS}^\prime = \hat{c}_{\kf \mathbf{n_Q}
,\up} \ket{\rm FS}$ with $\mathbf{n_Q}=\Q/Q$ a unit vector in the direction of $\Q$, then we see that Eq.~\eqref{eq:mol-wfn} corresponds to a special case of Eq.~\eqref{eq:pol-wfn},  where 
$\phi_{\Q;\k} = \alpha_{\p;\k\q}\, \delta_{\p+\q,\Q} \, \delta_{\q,\kf \mathbf{n_Q}}$ and $\phi_{\Q;\kf \mathbf{n_Q}} = \alpha_{\p;0} \, \delta_{\p, \Q-\kf \mathbf{n_Q}}$. 
Thus, the zero-momentum molecule state is contained in the $p=\kf$ polaron wave function and manifests as a minimum in the energy spectrum  
near the polaron-molecule transition, as shown in Fig.~\ref{fig:PolaronAndMoleculeHoleEdge}.
In particular, at this level of approximation, the sharp polaron-molecule transition corresponds to the point where the energy minima at $p=0$ and $p=\kf$ become degenerate, as first pointed out in Ref.~\cite{Mathy2011} (and also observed in later works~\cite{Trefzger2012,Edwards2013,Cui2020}).
This might suggest that the dimer simply corresponds to a polaron at finite momentum~\cite{Edwards2013,Cui2020}. 
However, the key point is that the dimer state has a residue that scales inversely with the volume~\footnote{Reinstating the volume $V$, we have $\phi_{\Q;\k} \to \phi_{\Q;\k}/\sqrt{V}$ due to the normalization $\sum_\k |\phi_{\Q;\k}|^2/V = 1$. Thus, the overlap with any of the non-interacting states scales with $1/\sqrt{V}$.} and thus the dimer is qualitatively different from the polaron in the \textit{thermodynamic limit} where the residue vanishes. 

Even though the molecule has a vanishing residue, it is still visible in the impurity spectral function 
because there is a continuum of states in Eq.~\eqref{eq:pol-wfn} involving a molecule and a hole excitation of the Fermi sea~\cite{Massignan2014}, as depicted in Fig.~\ref{fig:PolaronAndMoleculeHoleEdge}. The state in Eq.~\eqref{eq:mol-wfn} then defines the onset of this molecule-hole continuum since a hole excitation at the Fermi surface has the lowest energy. 
Thus, at the level of the single particle-hole approximation in Eq.~\eqref{eq:pol-wfn}, the spectral function near the transition $(1/\kf a)_0$ contains a well-defined polaron peak at low momenta and a visible molecule-hole continuum centered around $p=\kf$.  

In the exact spectral function, 
where one can in principle have an infinite number of particle-hole excitations, there is a continuum spanning all $p$ that extends all the way down to the ground-state energy (see Fig.~\ref{fig:PolaronAndMoleculeHoleEdge}) 
due to low-energy excitations at the Fermi surface. However, states with multiple particle-hole excitations have negligible spectral weight since their residue vanishes even faster with volume compared to the molecule-hole continuum. Indeed, the strong suppression of spectral weight above the polaron
has also been found in non-perturbative QMC calculations for unitarity-limited impurity-fermion interactions~\cite{Goulko2016}.
Thus, we expect the ansatz \eqref{eq:pol-wfn}, as well as \eqref{eq:polT}, to capture the main features of the spectral function. In particular, the double-minimum structure in Fig.~\ref{fig:PolaronAndMoleculeHoleEdge} should be observable in a real experiment. 
While additional particle-hole excitations will also couple the polaron and molecule minima in the spectrum, the phase space for such processes vanishes at the transition~\cite{Bruun2010} in the thermodynamic limit. 
Therefore the polaron-molecule transition is akin to the orthogonality catastrophe~\cite{Anderson1967} since it only formally exists in the limit where the number of majority $\up$ fermions tends to infinity.
Such a transition may resemble a crossover if the number of fermions considered is too small~\cite{Bour2015}.
    
\subsection{Free energy and contact}

\begin{figure}
    \centering
    \includegraphics[width=\linewidth]{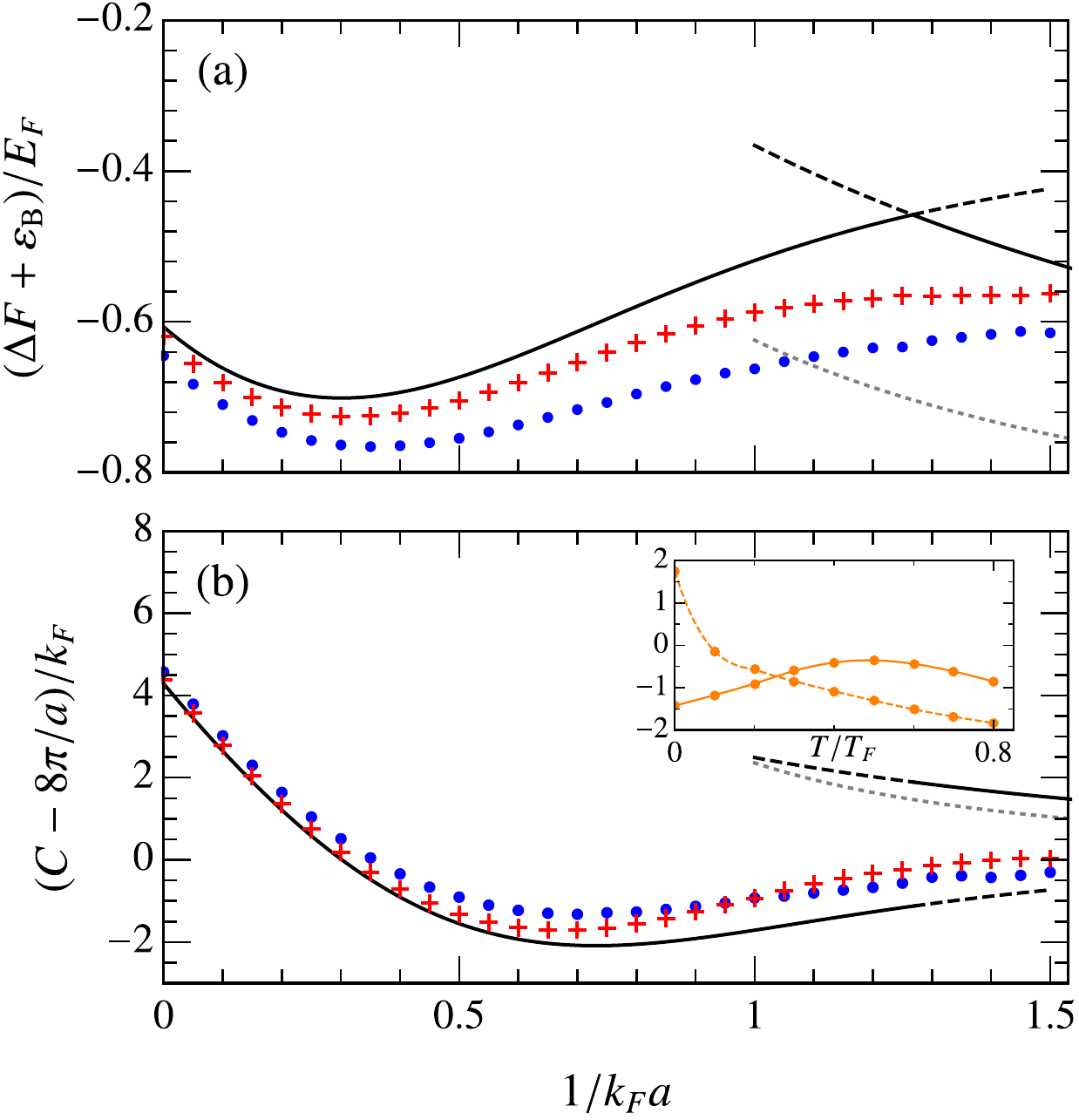}
    \caption{Impurity free energy (a) and contact (b) as a function of interaction strength around the polaron-molecule transition, $(1/k_F a)_0 \simeq 1.27$. In both cases, we have subtracted the leading contribution that arises from 
    the vacuum two-body bound state.  We show the $T=0$ results as solid (dashed) lines when the respective lines correspond to the ground (excited) states. The red crosses and blue circles are the results at $T/T_F = 0.1$ and 0.2, respectively, and we estimate the numerical error to be significantly smaller than the symbol size. The gray dotted lines correspond to the expression for a bosonic molecule weakly interacting with the Fermi sea 
    (see text). The inset in (b) shows the temperature dependence of the contact at $1/(k_F a) =0.5$ (solid) and $1.3$ (dashed), 
    where the lines are guides to the eye.}
    \label{fig:freecontact}
\end{figure}

We now turn to the fate of the polaron-molecule transition at finite temperature and how it is connected to the thermodynamic properties of the Fermi polaron. Using the relationship in Eq.~\eqref{eq:det-bal2} together with the sum rule \eqref{eq:sum2}, we obtain the following expression for the impurity free energy~\cite{Liu2020a,Liu2020b}
\begin{subequations}
\label{eq:free}
\begin{align}
    \label{eq:freea}
    e^{-\beta \Delta F} & = \frac{\int d\omega \sum_\p e^{-\beta \omega} A(\p,\omega)}{\sum_\p e^{-\beta \ep}} \\
    & = \left(\frac{2 \pi}{mT} \right)^{3/2} \sum_{\p, l} e^{-\beta E_\p^{(l)}} \big|\alpha_{\p;0}^{(l)}\big|^2\,.\label{eq:freeb}
\end{align}
\end{subequations}
This allows us to calculate the contact~\cite{Tan2008,Braaten2008}, which is defined by 
\begin{align}    \label{eq:contact}
    C = 4\pi m \left.\pdv{F}{(-1/a)}\right|_{T,\mu} = 4\pi m \left.\pdv{\Delta F}{(-1/a)}\right|_{T,\mu}\,.
\end{align}
The contact can also be extracted from the high-frequency tail~\cite{Braaten2010,Schneider2010} of the rf spectrum, since we have
\begin{align}
    I_{\rm ej}^{\rm rf}(\omega) &\to \frac{1}{4\pi^2 \sqrt{m}} \frac{C}{\omega^{3/2}}.
\end{align}
As long as $\omega\gg\eq$, the same relation holds for the Raman spectrum.

As illustrated in Fig.~\ref{fig:freecontact}(a), the impurity free energy calculated from our ansatz~\eqref{eq:polT} features a kink at the polaron-molecule transition in the zero-temperature limit, where the quasiparticle abruptly changes its character.
This agrees with the results for the impurity ground-state energy obtained from the variational wave functions \eqref{eq:pol-wfn} and \eqref{eq:mol-wfn}. Such a kink also exists when more particle-hole pair excitations are included~\cite{Prokofev2008,Mora2009,Punk2009,Combescot2009,Bruun2010,Vlietinck2013} although it then occurs at smaller $1/\kf a$, primarily
due to the lower energy of the molecule [see Fig.~\ref{fig:freecontact}(a)]. 
Note that we show $\Delta F+\eb$ to highlight the difference from the trivial part of the free energy originating from the two-body binding energy. We can understand the kink in the free energy from the behavior of the spectra shown in Fig.~\ref{fig:PolaronAndMoleculeHoleEdge}: When $1/\kf a<(1/\kf a)_0$, the free energy is set by the well-defined quasiparticle peak at zero momentum, while when $1/\kf a>(1/\kf a)_0$, the value of the free energy corresponds to the bottom of the molecule-hole continuum at $p=\kf$. As discussed above, the edge of this continuum extends to zero momentum when more particle-hole pairs are included; however the behavior of the free energy across the transition will still look qualitatively similar to Fig.~\ref{fig:freecontact}(a), and the transition can be characterized by the sudden loss of quasiparticle weight of the ground state.

\begin{figure}
	\centering
	\includegraphics[width=\linewidth]{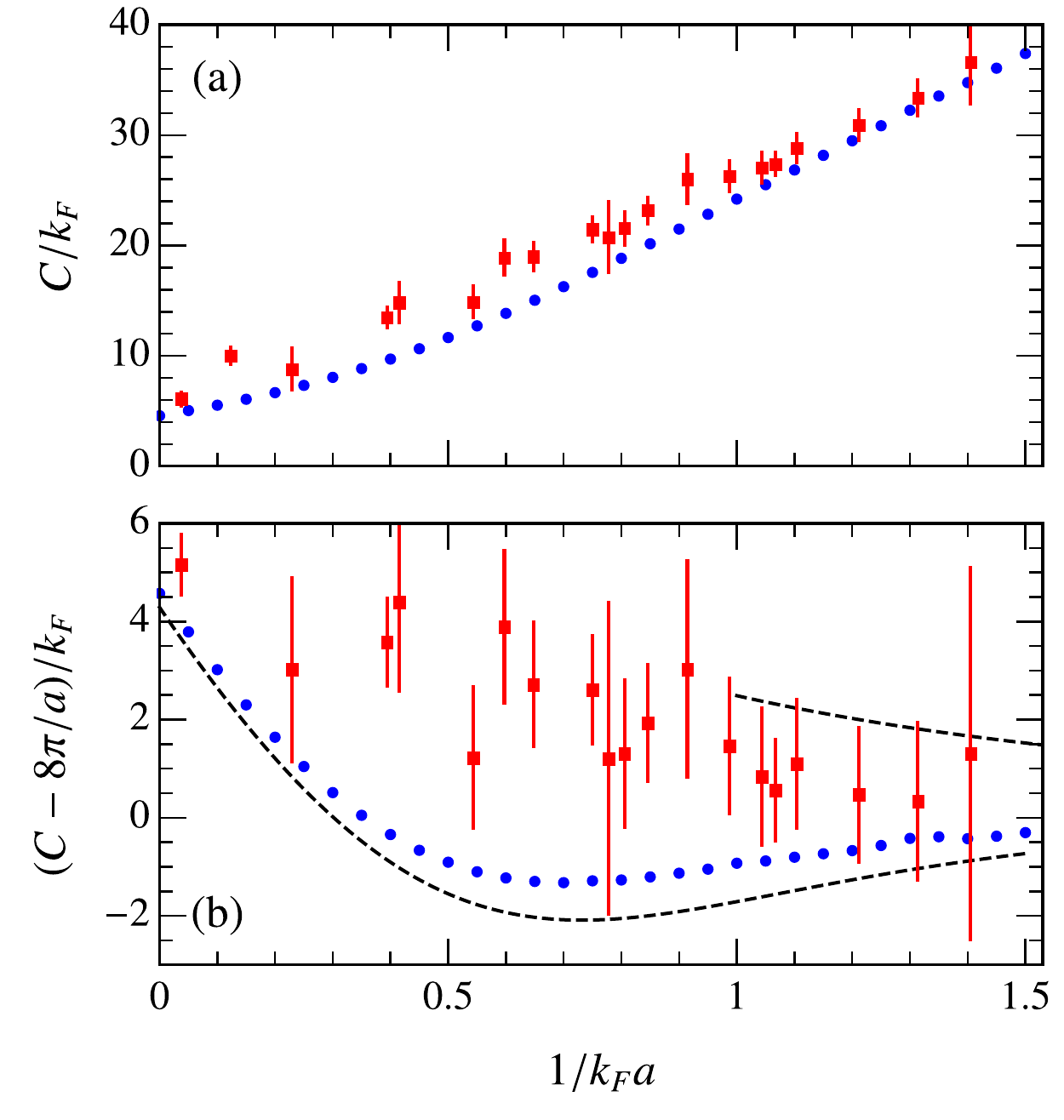}
	\caption{Contact measured in experiment (red squares)~\cite{Ness2020}, compared with our variational result at $T/T_F=0.2$ (blue circles), which is approximately the temperature in experiment. We make a direct comparison between our theoretical result and the experimental data in (a) and we subtract the contribution from the two-body bound state in (b). We also include the result at $T=0$ (dashed lines) for comparison.}
	\label{fig:contactraman}
\end{figure}

\begin{figure*}
	\centering
	\includegraphics[width=\linewidth]{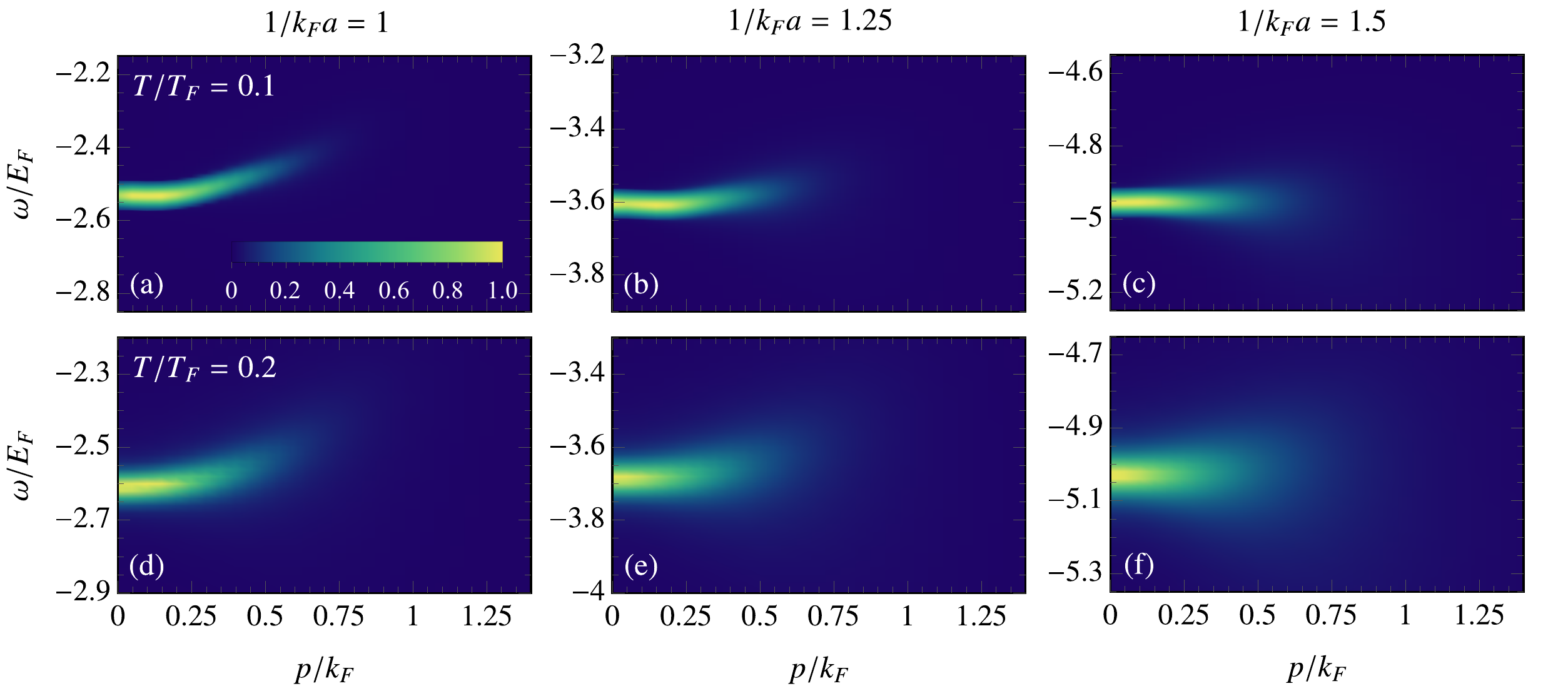}
	\caption{Occupied spectral function $e^{-\beta \omega} A(\p, \omega)$ close to the polaron-molecule transition at temperature $T/T_F=0.1$ (top row) and $0.2$ (bottom row). From left to right, the interactions in each column are $1/k_F a=1$, $1.25$ and $1.5$. 
	In all panels we apply a Gaussian broadening with standard deviation $0.02 E_F$, and we normalize the spectrum to the maximum value in each panel.}
	\label{fig:occuspec}
\end{figure*}

\begin{figure*}
	\centering
	\includegraphics[width=\linewidth]{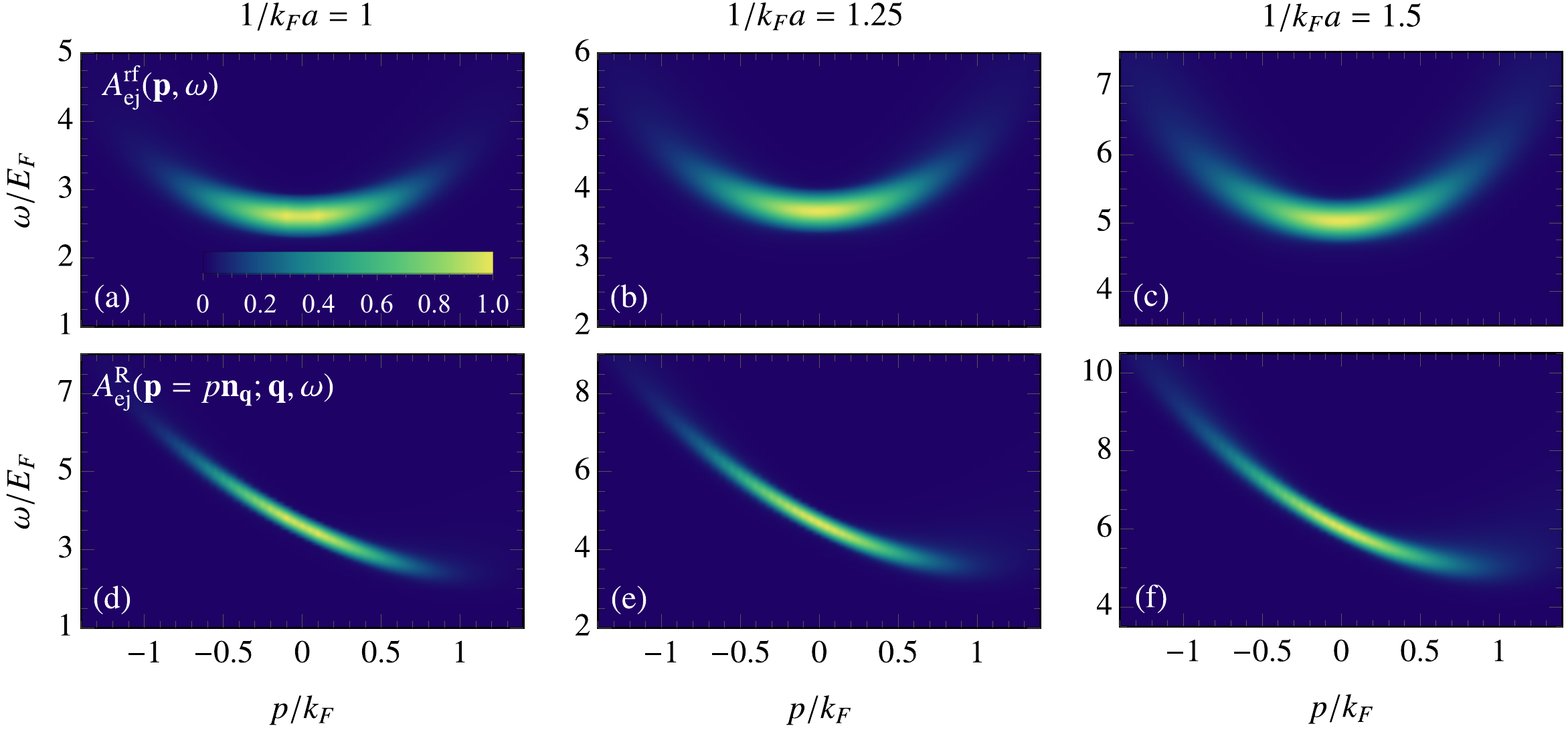}
	\caption{Impurity ejection spectra close to the polaron-molecule transition as a function of  momentum $p$ along the direction of $\q$. We have temperature $T/T_F = 0.2$ and, from left to right, the interactions in each column are $1/k_F a=1$, $1.25$ and $1.5$. In the top row, we show rf ejection spectroscopy, while in the bottom row, we show Raman ejection spectroscopy with $q= k_F$. Across the transition, we observe spectral weight spreading to the molecule-hole continuum at $\p=\pm k_F \mathbf{n_q}$. In all panels we apply a Gaussian broadening with standard deviation $0.15 E_F$, and we normalize the spectrum to the maximum value in each panel.}
	\label{fig:ejection}
\end{figure*}

\begin{figure*}
	\centering
	\includegraphics[width=\linewidth]{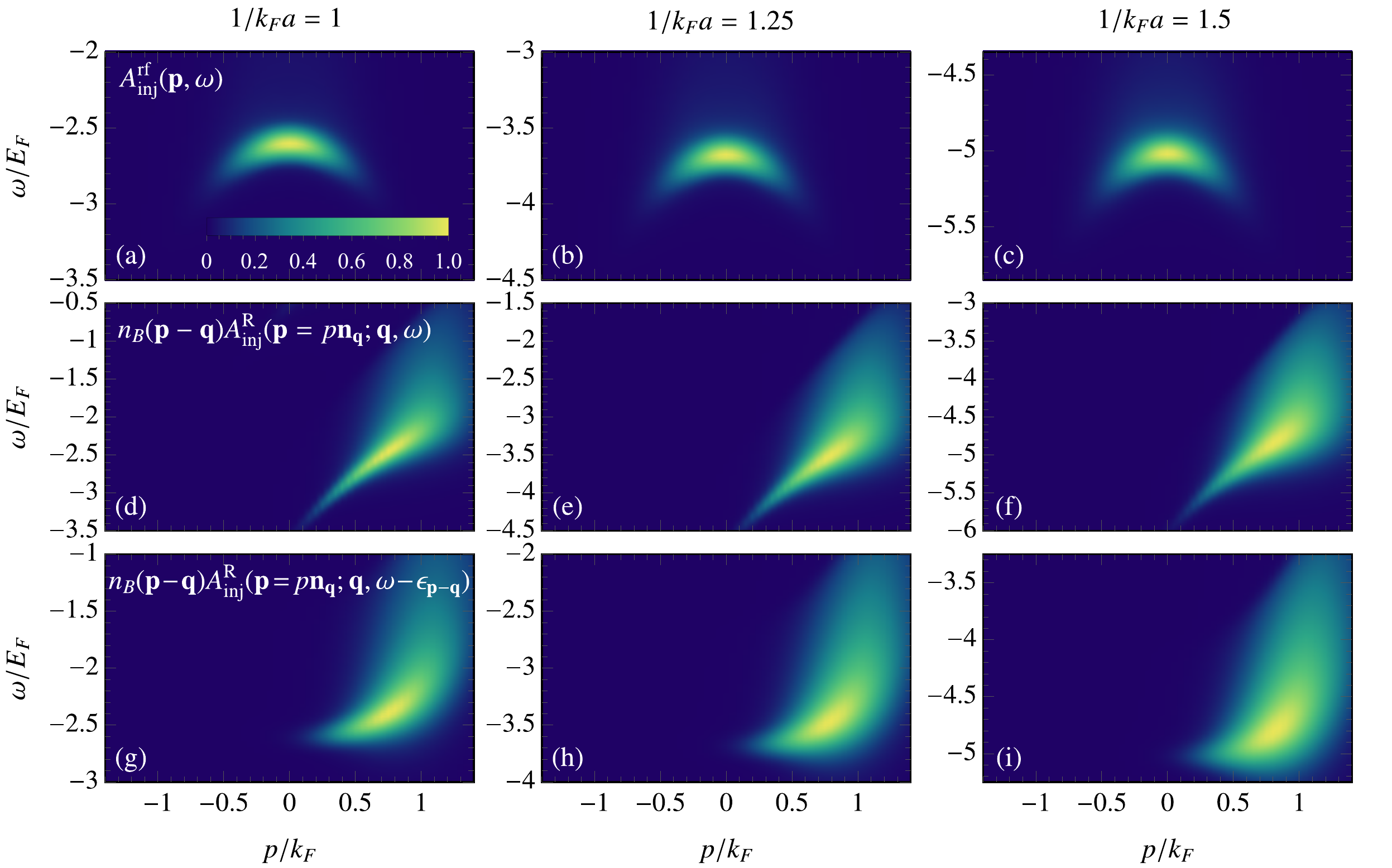}
	\caption{Impurity injection spectra close to the polaron-molecule transition at temperature $T/T_F = 0.2$. From left to right, the interactions in each column are $1/k_Fa=1$, $1.25$ and $1.5$. In the top row, we show rf injection spectra, where we weight the spectrum by the Boltzmann distribution $n_{\rm B}(\p)$. In the middle and bottom row, we show Raman injection spectroscopy with $q= k_F$ and $\p=p \mathbf{n_q}$, where we weight the spectrum by $n_{\rm B}(\p - \q)$. The choice of two-photon wave vector enables Raman spectroscopy to sensitively probe the molecule-hole continuum contribution at $\p = k_F \mathbf{n_q}$. In the bottom row, we counter the frequency shift caused by the Raman spectroscopy to observe the molecule-hole continuum onset lowering in energy relative to the attractive polaron at $\p=0$ across the polaron-molecule transition. In all panels we apply a Gaussian broadening with standard deviation $0.06 E_F$, and we normalize the spectrum to the maximum value in each panel.}
	\label{fig:injection}
\end{figure*}

At finite temperature, Fig.~\ref{fig:freecontact}(a) clearly illustrates how the kink in the free energy is replaced by a smooth crossover, where it becomes difficult to distinguish the polaron-molecule transition already when $T\gtrsim0.1T_F$. This behavior arises from the thermal population of excited states in the impurity spectral function, as well as from the thermal smearing of the $\up$ Fermi surface which modifies the spectral function itself. The latter effect is expected to be relevant for $T\gtrsim0.1T_F$~\cite{Tajima2018} but has not previously been included in variational studies of the polaron-molecule transition at finite temperature~\cite{Ness2020,Cui2020}.
We also note that the magnitude of the impurity free energy initially increases with temperature at low temperatures $T\ll T_F$, which is in contrast to the behavior at high temperatures where we have $\Delta F \to 0$.
This low-temperature feature can be traced back to the fact that the interacting spectral function in Eq.~\eqref{eq:freea} has a larger density of states at low energies compared to the non-interacting case (see Sec.~\ref{sec:spectra}), and thus $\Delta F$ is lowered as these states become thermally populated. Since contributions to the free energy at finite temperature are dominated by regimes of significant spectral weight --- see Eq.~\eqref{eq:free} --- the finite-temperature free energy shown in Fig.~\ref{fig:freecontact}(a) is unlikely to strongly depend on the number of particle-hole pairs included in the variational ansatz.

We now argue that the temperature dependence of the contact, shown in Fig.~\ref{fig:freecontact}(b), provides a clear signature of the underlying zero-temperature single-impurity transition. As in the case of the free energy, we have subtracted the (positive) contribution to the contact from the two-body bound state. At $T=0$, the kink in the free energy leads to an abrupt jump in the contact across the transition. This discontinuity was also observed in Ref.~\cite{Punk2009}, where the $T=0$ contact was first obtained. At low but finite temperature, we find that the contact initially increases on the polaron side of the transition since temperature populates the molecule-hole continuum which has a higher contact than the polaron itself~\cite{Liu2020a,Liu2020b}. On the other hand, we expect the contact to decrease monotonically on the molecule side of the transition since both the molecular excited states and the polaron have a smaller contact than the ground-state dimer. We therefore predict a strong qualitative difference in the behavior of the contact as a function of temperature on the polaron and the molecule side of the transition. Indeed, this is clearly observed in our numerical results shown in Fig.~\ref{fig:freecontact}(b). Unlike the kink in the free energy, this non-monotonic behavior should be discernible at typical experimental temperatures $T/T_F\sim 0.1$--0.3. 

In the regime where $1/\kf a\gtrsim1$ and $T \lesssim \eb$, we can approximately treat the system as a bosonic molecule that is weakly interacting with the Fermi sea. At zero temperature, this gives $\Delta F+\eb\to-E_F+nU_{BF}$, where the Bose-Fermi interaction $U_{BF}=3\pi a_{ad}/m$, which involves the atom-dimer scattering length $a_{ad}=1.18a$~\cite{Skorniakov1957}. This has been found to well approximate the ground-state energy of the dressed molecule~\cite{Combescot2009}. The corresponding line is shown in Fig.~\ref{fig:freecontact}(a), and we see that it lies somewhat below that obtained from the variational wave function in Eq.~\eqref{eq:mol-wfn}, highlighting the fact that Eq.~\eqref{eq:mol-wfn} overestimates the strength of the atom-dimer interactions. This can be cured by including more particle-hole pairs in our variational ansatz~\cite{Mora2009,Punk2009,Combescot2009}, leading to a correction to the position of the polaron-molecule transition. Likewise, the contact for a molecule interacting with majority fermions is given by $C\simeq 8\pi/a+2\kf (\kf a)^2 a_\mathrm{ad}/a$. In this case, the corresponding line in Fig.~\ref{fig:freecontact}(b) lies quite close to our variational result, and therefore it is likely that corrections due to the inclusion of multiple particle-hole pairs will be small.

Experimentally, the contact has been extracted from the large frequency tail in rf ejection spectra for the case of unitarity limited impurity-fermion interactions~\cite{Yan2019}, and more recently the contact was extracted over a larger range of interactions using the high-frequency tail in Raman ejection spectra~\cite{Ness2020}. We have previously shown that the temperature dependence at unitarity observed in Ref.~\cite{Yan2019} can be well reproduced within our variational approach~\cite{Liu2020a,Liu2020b}. Figure~\ref{fig:contactraman} shows a comparison between the results of Ref.~\cite{Ness2020} and our variational ansatz at $T/T_F = 0.2$ (the approximate temperature in experiment). On the absolute scale shown in panel (a) we find a good agreement. However, in panel (b) we see that once we subtract the two-body contribution that dominates already when $1/\kf a\gtrsim0.2$, the results deviate substantially. This difference could, at least in part, be due to the trap averaging in experiment~\cite{Ness2020}, the finite density of impurities, and possibly non-negligible effective range corrections in the two-body scattering phase shift in $^{40}$K. 

\subsection{Impurity spectra in the vicinity of the transition}
\label{sec:spectra}

We now address how signatures of the polaron-molecule transition can be observed via the spectroscopic techniques detailed in Sec.~\ref{sec:probes}. The initial state of the interacting impurity in a Fermi sea is encoded in the occupied spectral function, defined by $e^{-\beta \omega} A(\p, \omega)$, which in turn controls the thermodynamic properties via its relation to the impurity free energy, Eq.~\eqref{eq:freea}. In Fig.~\ref{fig:occuspec} we show our calculated occupied spectral function at temperatures $T/T_F=0.1$ and $0.2$ around the polaron-molecule transition. 
The molecule-hole continuum centered at $p=\kf$ (see Fig.~\ref{fig:PolaronAndMoleculeHoleEdge}) is barely visible due to its small spectral weight, but we clearly see that the impurity dispersion flattens with increasing attraction, which is 
a sign that the molecule-hole continuum lowers relative to the polaron. 
At lower temperature, we also observe a bending down of the polaron dispersion as it approaches $p=\kf$ where it starts coupling to the molecule-hole continuum. In order to better expose the polaron-molecule transition, we require a spectroscopic protocol that enhances the continuum around  $p=\kf$. Thus we will now investigate the possibility of using Raman spectroscopy. 

In Fig.~\ref{fig:ejection} we compare the rf and Raman ejection spectra across the transition. These can be related to the spectral function using Eqs.~\eqref{eq:Ainj}, \eqref{eq:ejrf}, \eqref{eq:spec-rel} and \eqref{eq:det-bal1}:
\begin{subequations}
\begin{align}
    A_\mathrm{ej}^\mathrm{rf}(\p,\omega) &=e^{\beta\,\Delta F}e^{\beta\omega}n_B(\p)A(\p,-\omega+\ep), \\
    A_\mathrm{ej}^\mathrm{R}(\p;\q,\omega) &=e^{\beta\,\Delta F}e^{\beta\omega}n_B(\p-\q)A(\p,-\omega+\epsilon_{\p-\q}).
\end{align}
\end{subequations}
In the case of Raman spectroscopy, we choose the two-photon momentum to be at $q = k_F$ in order to add weight to the molecule-hole continuum. The spectrum then depends on the direction of the final impurity momentum with respect to $\q$, and we choose to plot this in the direction parallel (positive $p$) or antiparallel (negative $p$) to the Raman momentum. On the molecule side of the transition, both rf and Raman spectra show slightly increased spectral weight at $p= \pm k_F$, which signifies an increased contribution from the molecule-hole continuum. However, it appears that neither spectroscopic technique shows marked differences across the transition.

Finally, in Fig.~\ref{fig:injection} we contrast rf and Raman injection spectra in the vicinity of the transition. Here we weight the spectra by the distribution function of the initial momentum state (i.e., $n_B(\p)$ for rf and $n_B(\p - \q)$ for Raman) to show the contributions to the total spectral function sums in Eq.~\eqref{eq:Iinj} and Eq.~\eqref{eq:Iinj-q}, respectively. Similar to the occupied spectral function, rf injection spectroscopy is dominated by low momentum attractive polaron states, with the molecule-hole continuum holding little spectral weight, and therefore the spectrum exhibits no marked changes across the transition. By contrast, the additional momentum parameter, $\q$, in Raman injection spectroscopy overcomes these limitations. In particular, one can take advantage of the weighting function, $n_B(\p - \q)$, to add significant spectral weight around $\p = \q$. With $q = k_F$, we find that the molecule-hole continuum around $\p=\kf\mathbf{n_q}$ dominates the spectrum. Furthermore, by subtracting the frequency shift caused by the Raman spectroscopy, $\epsilon_{\p- \q}$ [see Eq.~\eqref{eq:spec-rel}], we see clear evidence of the molecule-hole continuum onset lowering in energy relative to the energy of the $\p=0$ attractive polaron. We thus conclude that Raman injection spectroscopy is particularly useful close to the transition.

From an experimental perspective, our results in Figs~\ref{fig:occuspec}--\ref{fig:injection} suggest that momentum-resolved Raman injection spectroscopy with $q = k_F$ is well suited to probe the polaron-molecule transition. As shown in the bottom row of Fig.~\ref{fig:injection}, post-processing of the spectra, where one subtracts the frequency shift caused by Raman spectroscopy, $\epsilon_{\p- \q}$, would provide clean access to the energy of the molecule-hole continuum onset (at $\p=k_F \mathbf{n_q}$) for experimentally accessible temperatures.
To date, momentum-resolved spectroscopy has been achieved with rf ejection spectroscopy of Fermi polarons in two dimensions~\cite{Koschorreck2012}, and with rf injection spectroscopy of zero-momentum Bose polarons~\cite{Jorgensen2016} in three dimensions through transferring the impurities from a Bose-Einstein condensate. Thus, it appears feasible that momentum-resolved Raman injection spectroscopy will be realized in future experiments. We also note that the momentum-averaged Raman spectrum with $q=\kf$ will be dominated by regions of largest spectral weight, and therefore by the molecule-hole continuum.

\begin{figure}
	\centering
	\includegraphics[width=\linewidth]{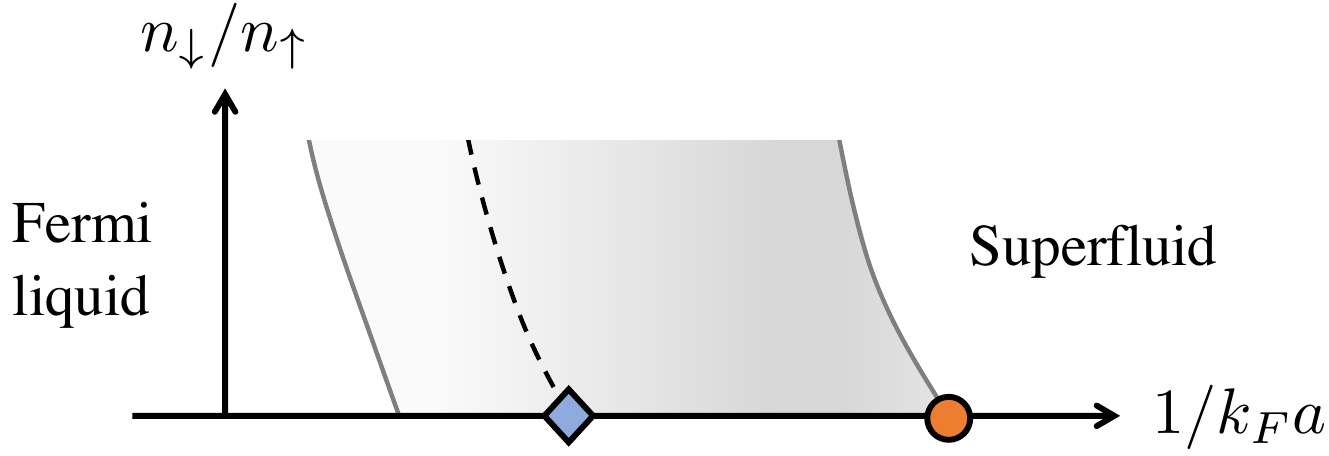}
	\caption{Schematic illustration of the zero-temperature phase diagram in the highly spin-imbalanced regime. The blue diamond denotes the polaron-molecule transition in the single-impurity limit $n_\down \to 0$. The associated 
	continuous 
	transition (dashed line) between superfluid and Fermi liquid phases is preempted by a first-order transition, resulting in a region of phase separation (shaded area) that terminates in a tricritical point (orange circle).}
	\label{fig:phaseT0}
\end{figure}

\section{Spin-imbalanced phase diagram}
\label{sec:phase}

The behavior of the Fermi polaron has implications for the phase diagram of the spin-imbalanced Fermi gas since it determines the possible phases in the limit of extreme polarization, where $n_\down/n_\up \to 0$.  
In particular, a polaron-molecule transition in the single-impurity limit implies the existence of a phase transition between a normal Fermi liquid and a paired-fermion superfluid
at finite impurity density, as illustrated in Fig.~\ref{fig:phaseT0}. If the single-impurity transition is thermodynamically  stable --- in the sense that it applies to a global minimum of the Fermi system's free energy --- then it would correspond to a continuous (second-order) phase transition. 
However, BCS mean-field theory~\cite{Radzihovsky2010} and variational QMC~\cite{Pilati2008} calculations predict that such a transition is preempted by a \textit{first-order} phase transition at zero temperature, where the system spatially separates into domains of superfluid and Fermi liquid phases (see Fig.~\ref{fig:phaseT0}). Such phase separation has also been observed in cold-atom experiments~\cite{Partridge2006,Shin2006}.

The first-order transition at positive scattering length can be understood as arising from interactions between the dressed impurity quasiparticles.  
Specifically, it occurs when there is an effective attraction between the dressed molecules or dimers. To see this, consider the effective Ginzburg-Landau free energy for the paired superfluid at zero temperature and large spin imbalance
\begin{align} \label{eq:GL}
    \Omega = -\mu_d |\varphi|^2 + \frac{u}{2}|\varphi|^4 ,
\end{align}
where the order parameter $|\varphi|^2$ corresponds to the density of $\up\down$ dressed dimers, while $\mu_d$ is the dimer chemical potential and $u$ is the strength of the interactions between dressed dimers. In the limit of a single impurity, one ignores the quartic interaction term and only considers the behavior of the quadratic (one-body) term. 
For $\mu_d <0$, the polaron is favored over the dressed dimer and there is no superfluid phase ($\varphi = 0$). Thus, the polaron-molecule transition corresponds to the point where $\mu_d$ first becomes zero, such that the dressed dimer energy is equal to the polaron energy. 

Beyond the single-impurity limit, the condition $\mu_d=0$ in Eq.~\eqref{eq:GL} corresponds to a second-order phase transition 
provided we have repulsive dimer-dimer interactions $u>0$. This is the case for strong attraction, $1/\kf a \gg 1$, where one has a superfluid of tightly bound repulsive $\up\down$ pairs that can coexist with any excess $\up$ fermions, as depicted in Fig.~\ref{fig:phaseT0}.  For $\mu_d>0$, we have a finite density of bosonic pairs given by $n_\down=\mu_d/u$, so the condition $\mu_d = 0$ describes a second-order phase boundary at $n_\down =0$ where the superfluid density smoothly goes to zero.
For decreasing $1/\kf a$ along this phase boundary, the onset of phase separation occurs at a tricritical point (orange circle in Fig.~\ref{fig:phaseT0}), where the effective interactions vanish and we have $\mu_d = u = 0$. We discuss in Sec.~\ref{sec:tricrit} 
how this point can be understood as arising from the competition between direct and induced boson-boson interactions in a Bose-Fermi mixture.

Continuing along the phase boundary at $n_\down =0$, immediately to the left of the tricritical point in Fig.~\ref{fig:phaseT0} the effective dimer-dimer interactions are attractive. Consequently, the polaron-molecule transition (after which $\mu_d<0$)
marks the appearance of a \textit{local} minimum in the free energy at $\varphi = 0$, since $u<0$ means that Eq.~\eqref{eq:GL} is unbounded from below and we must consider higher order terms in the density $|\varphi|^2$.
The polaron-molecule transition therefore corresponds to the end of
a spinodal line (dashed line in Fig.~\ref{fig:phaseT0}), beyond which the Fermi liquid appears as a metastable state.
According to QMC calculations~\cite{Pilati2008}, the Fermi liquid is finally stabilized at weaker attraction (closer to unitarity), as illustrated in Fig.~\ref{fig:phaseT0}.
However, we note that it has also been conjectured~\cite{Chevy2010,Frank2018} that the polaron-molecule transition coincides with the onset of phase separation, which would then require $u$ to change sign at the transition. Such a scenario appears unlikely given that the induced interactions between dimers are expected to be strongly attractive in this regime (see Sec.~\ref{sec:tricrit}).
A full description of the phase separated state requires an analysis beyond the large imbalance limit of Eq.~\eqref{eq:GL}, since the superfluid regions may only be weakly spin-polarized or even unpolarized~\cite{Radzihovsky2010,Chevy2010}.

Before turning to the tricritical point, let us conclude this section by examining the situation where there is no sharp single-impurity transition at zero temperature. Such a smooth crossover 
is expected to occur in the case of the Bose polaron~\cite{Rath2013}, where the impurity is immersed in a Bose Einstein condensate rather than a Fermi gas. The absence of a single-impurity transition implies that there is no symmetry-breaking continuous transition in a dilute gas of impurities, in contrast to the superfluid transition discussed above. This makes sense in the case of the Bose polaron since the statistics of the impurity is unchanged if it binds a boson from the medium, and thus the impurity quasiparticle should not fundamentally change its character.
However, this does not preclude the possibility of a first-order phase transition between states of the same symmetry but differing densities.

\subsection{Tricritical point at zero temperature} 
\label{sec:tricrit}

\begin{figure}
	\centering
	\includegraphics[width=.8\linewidth]{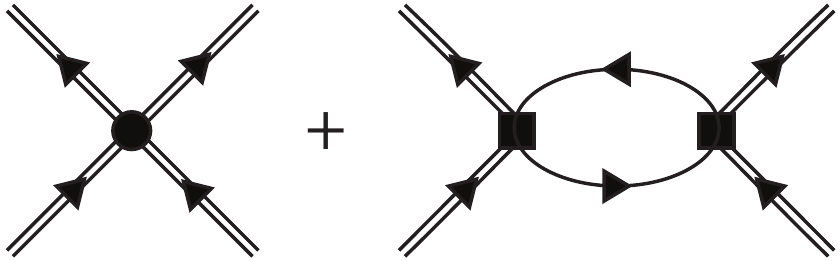}
	\caption{Effective interaction between bosons in the effective Bose-Fermi mixture at $1/\kf a\gtrsim1$. The boson-boson and boson-fermion interactions, $U_{BB}$ and $U_{BF}$, are drawn as a circle and squares, respectively. The double lines correspond to bosons and the single lines to fermions.}
	\label{fig:feynman}
\end{figure}

The tricritical point in the spin-imbalanced phase diagram corresponds to the point where the superfluid-normal phase transition changes from first- to second-order with increasing $1/\kf a$.
To determine the tricritical point at zero temperature, we consider the highly polarized limit in the case of strong attractive interactions where $1/\kf a\gtrsim1$. In this limit, we can approximate the gas as a repulsive Bose-Fermi mixture consisting of bosonic dressed dimers and excess majority fermions. As argued above, the tricritical point occurs when the effective boson-boson interaction vanishes. As illustrated in Fig.~\ref{fig:feynman}, there are two contributions to this effective interaction, corresponding to the direct boson-boson interaction $U_{BB}$ and the induced interaction mediated by the gas of excess fermions $U_\mathrm{ind}$:
\begin{align}
    u=U_{BB}+U_\mathrm{ind}.\label{eq:u}
\end{align}
The induced interaction is a second order process in the boson-fermion interaction $U_{BF}$, and takes the form
\begin{align}
    U_\mathrm{ind}=\Pi(0)U_{BF}^2,
\end{align}
where $\Pi(0)=-m\kf/2\pi^2$ is the static Lindhard function of excess fermions in the long wavelength limit. Note that whereas the direct interaction is repulsive, the induced interaction is attractive.

To proceed, we note that the boson-boson and boson-fermion interactions are
\begin{align} 
 U_{BB} & = \frac{2\pi a_{dd}}{m}\,, \qquad
 U_{BF} = \frac{3\pi a_{ad}}{m} \, ,\label{eq:Us}
\end{align}
in terms of the dimer-dimer and atom-dimer scattering lengths $a_{dd}$ and $a_{ad}$, respectively. Here we have used the fact that the boson-boson and boson-fermion reduced masses are $m$ and $2m/3$, respectively. Setting $u=0$ and using Eqs.~\eqref{eq:u}-\eqref{eq:Us} we thus find that the tricritical point occurs when
\begin{align}
 \left(\frac{1}{k_Fa}\right)_\mathrm{tcp} & =
 \frac{9}{4 \pi a} \frac{a_{ad}^2}{a_{dd}}.
    \label{eq:tcp}
\end{align}
Using the known values $a_{dd}=0.6 a$ \cite{Petrov2004} and $a_{ad} = 1.18 a$~\cite{Skorniakov1957}, we obtain
$(1/k_{F}a)_\mathrm{tcp} \simeq 1.7$. Remarkably, this value coincides with the result of QMC~\cite{Pilati2008}, contrary to what was claimed
in Ref.~\cite{Punk2009}. Note that the BCS
mean-field result of $(1/k_{F}a)_{\rm tcp} \simeq
1.88$~\cite{Parish2007a} is also quite close to the QMC result. This is because the upper critical dimension of a tricritical point is three and thus we expect mean-field theory to provide a reasonable description.

Equation~\eqref{eq:u} also enables us to evaluate the effective interaction in general in the regime $1/\kf a\gtrsim1$ where the highly polarized system is well described as a Bose-Fermi mixture. This yields
\begin{align}
    u=U_{BB}\left[1-\frac{\kf a}{(\kf a)_\mathrm{tcp}}\right].
\end{align}
In particular, at the polaron-molecule transition we find that the effective boson-boson interaction is $u\sim -U_{BB}$. Although this estimate is, strictly speaking, on the border of its regime of validity, the large negative value of $u$ indicates that it is unlikely that the onset of phase separation coincides with the single-impurity transition.

The criterion in Eq.~\eqref{eq:tcp} can also be obtained from a stability analysis of the Bose-Fermi mixture as discussed in Ref.~\cite{Chevy2010}. Here one considers the energy density~\cite{Viverit2000}:
\begin{align}
E & = \frac{3}{5} \frac{(6\pi^2\delta
n)^{2/3}}{2m} \delta n + U_{BF}\, n_{\downarrow} \delta n  +
\frac{1}{2} U_{BB} \,n_\downarrow^2,
\end{align}
where the excess fermion density $\delta n = n_\uparrow -n_\downarrow$. The system will become linearly unstable to phase
separation when the compressibility matrix ceases to be positive definite, \textit{i.e.}, when
\begin{align}\label{eq:mu_cond}
 \frac{\partial \mu_F}{\partial (\delta n)} \frac{\partial \mu_B}{\partial
 n_{\downarrow}} \leq & \
 \frac{\partial \mu_F}{\partial n_{\downarrow}} \frac{\partial \mu_B}{\partial (\delta n)},
\end{align}
where $\mu_F = \partial E/ \partial (\delta n)$ and $\mu_B =
\partial E/ \partial n_\downarrow$.
Taking the limit $n_\downarrow \rightarrow 0$ then yields Eq.~\eqref{eq:tcp}.

\subsection{Effective range corrections}
Thus far, we have discussed the scenario of a broad Feshbach resonance, where the two-body physics is determined by a single parameter, the $s$-wave scattering length $a$. It is interesting to consider what happens in the case of a narrow Feshbach resonance~\cite{Chin2010}. There, the momentum-dependent two-body scattering phase shift is instead determined by $k\cot\delta=-1/a-R^*k^2$, where the additional range parameter $R^*$ is inversely proportional to the resonance width~\cite{Petrov2004narrow} and we have $k_FR^*\gtrsim1$. Variational calculations~\cite{Massignan2012} have shown that the polaron-molecule transition occurs at $1/k_Fa\simeq -k_FR^*/2$ for $k_FR^*\gg1$, i.e., it moves to small negative scattering lengths. On the other hand, the tricritical point can still be obtained from Eq.~\eqref{eq:tcp}: Using the asymptotic expressions $a_{ad}\simeq 4a/3$ and $a_{dd}\simeq a\sqrt{a/R^*}/2$ valid when $R^*\gg a$~\cite{Levinsen2011}, we find that $(1/k_Fa)_{\rm tcp}\simeq (8/\pi)^2k_FR^*$, i.e., the tricritical point moves instead to small positive scattering lengths. Therefore, for a narrow Feshbach resonance, we expect the unitarity regime of the phase diagram to be dominated by phase separation.

\subsection{Effect of finite temperature} 
\begin{figure}
	\centering
	\includegraphics[width=\linewidth]{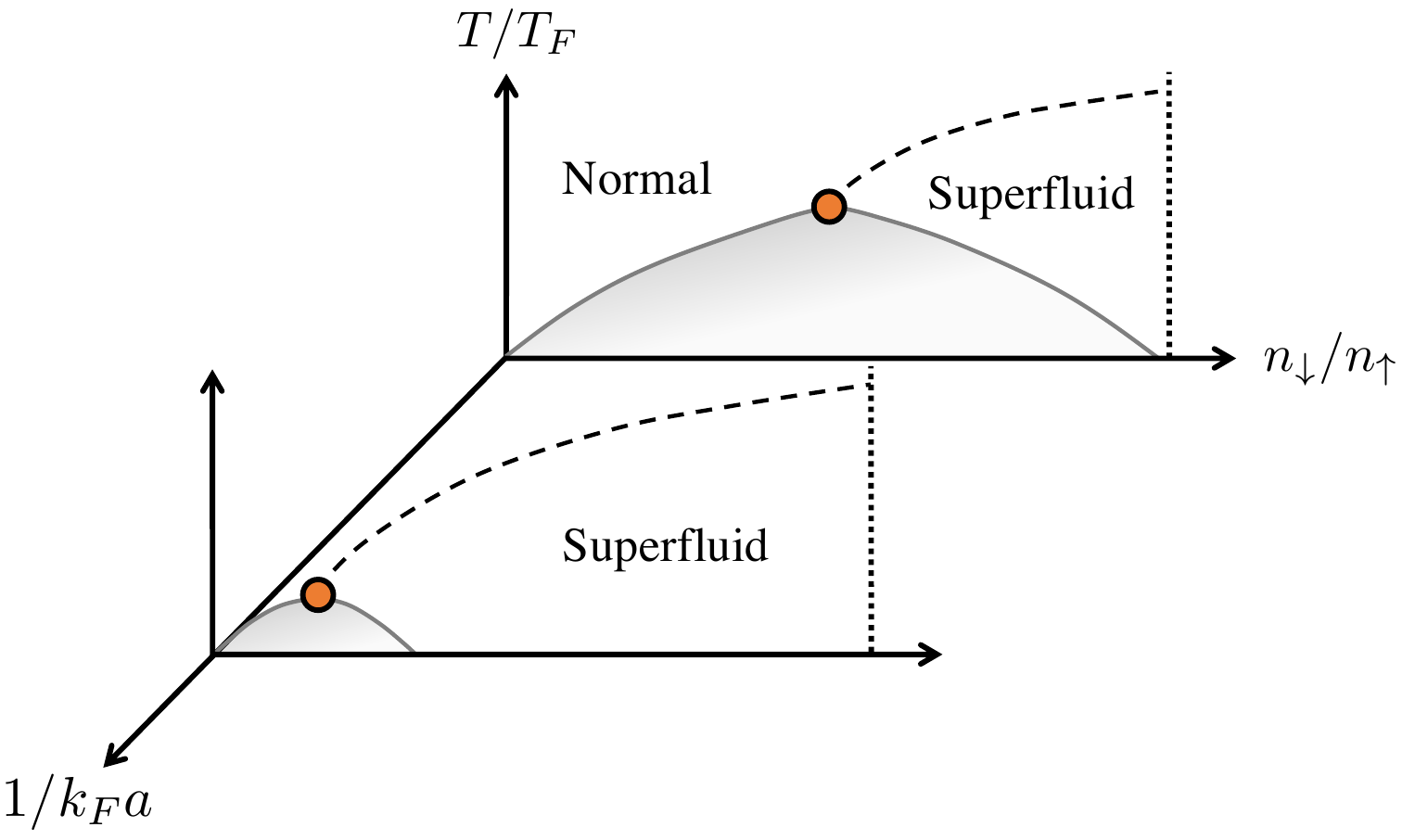}
	\caption{Schematic phase diagram at finite temperature for two values of $1/k_Fa$ near the $T=0$ polaron-molecule transition.  The first-order phase transition and accompanying region of phase separation (shaded area) becomes a continuous transition (dashed line) between normal and superfluid phases for temperatures above a tricritical point (orange circle). The dashed vertical lines denote the unpolarized system $n_\down/n_\up = 1$.}
	\label{fig:phase-finiteT}
\end{figure}

We now turn to the effect of temperature on the spin-imbalanced phase diagram and how this is related to the finite-temperature behavior of the Fermi polaron. Theoretically, it has been shown that there is a line of tricritical points at finite temperature that terminates at the zero-temperature tricritical point in Fig.~\ref{fig:phaseT0}~\cite{Parish2007a}. 
Each of these tricritical points marks the point where a second-order transition between superfluid and normal phases becomes first order, resulting in a dome of phase separation, as shown in Fig.~\ref{fig:phase-finiteT}. 
This finite-temperature phase diagram has also been confirmed experimentally at unitarity~\cite{Shin2008a,Navon2010}.

In the highly imbalanced limit $n_\down/n_\up \ll 1$, 
Fig.~\ref{fig:phase-finiteT} shows that increasing the impurity density $n_\down$ will eventually  result in phase separation for 
temperatures below the tricritical point, when the interaction $1/k_{F}a < (1/k_{F}a)_{\rm tcp}$. 
Therefore, a finite impurity density does not necessarily lead to coexistence between polarons and molecules near 
$(1/k_{F}a)_{0}$, in contrast to what has previously been claimed~\cite{Ness2020,Cui2020}. 
To obtain a smooth crossover between polarons and molecules, we require $T/T_F \gtrsim (n_\down/n_\up)^{2/3}$ such that the impurities form a classical Boltzmann gas and we can neglect any interactions (induced or otherwise) between impurities. 
This condition is also consistent with the phase diagram in Fig.~\ref{fig:phase-finiteT}, where we see that there is a uniform normal phase at sufficiently small $n_\down$ even when the temperature is below the tricritical point.
The recent Raman spectroscopy measurements~\cite{Ness2020} on the Fermi polaron appear to be in the regime $T/T_F \approx (n_\down/n_\up)^{2/3}$ and at temperatures that are possibly even above the tricritical point~\cite{Parish2007a}, 
thus leading to the observed polaron-molecule crossover. 

\section{Conclusion}
\label{sec:conc}
To conclude, we have investigated the thermodynamic and spectral signatures of the polaron-molecule transition in a Fermi gas. In particular, we have calculated the impurity free energy and contact across the transition, exposing how the transition is replaced by a smooth crossover at finite temperature. We have found that the contact has a characteristic non-monotonic temperature dependence on the polaron side of the transition which is absent on the molecule side, thus providing a clear and potentially observable signature of the underlying zero-temperature transition. We have furthermore discussed how the structure of the spectral function near the transition can be probed with Raman injection spectroscopy where the photon momentum is set to match the Fermi momentum, thus allowing the molecule-hole continuum to be clearly imaged.
 
For a finite density of impurities, we have argued that descriptions based on the behavior of a single impurity are only valid when $T \gg n_\down^{2/3}/m$, where we can neglect any correlations between impurities.
On the other hand, at sufficiently low temperatures, 
increasing the impurity density
leads to phase separation due to the strong induced attraction between dressed dimers mediated by excess fermions. 
Approximating the system as a Bose-Fermi mixture of dressed dimers and excess fermions, we have determined the tricritical point at $T=0$ to be at $1/\kf a\simeq1.7$ in agreement with the result of QMC~\cite{Pilati2008} (see also the stability analysis in Ref.~\cite{Chevy2010}). 

Our formalism can naturally be extended to the multitude of other impurity problems being actively pursued in the field of quantum gases. Of particular interest is the Bose polaron, where the statistics of the bosonic medium means that the single-impurity transition is replaced by a crossover even at zero temperature~\cite{Rath2013}. In this case there is no continuous (symmetry breaking) transition in the limit of vanishing impurity density, but the system may still feature a first-order transition depending on the precise details such as the statistics of the impurity.

Another promising direction of research is the Fermi polaron in two spatial dimensions. Here, a variational ansatz for the polaron with a single particle-hole excitation does not contain a polaron-molecule transition~\cite{Zollner2011}, and instead it is necessary to consider impurity dressing by two particle-hole pairs in order to reveal the transition~\cite{Parish2011}. While the investigation of the transition using our finite-temperature formalism therefore poses a technical challenge, the $T=0$ ground-state energy using such an ansatz has already been calculated~\cite{Parish2013}. Such a theory could potentially guide experiments on two-dimensional (2D) Fermi polarons~\cite{Koschorreck2012,Oppong2019} to reveal the transition.

Finally, 2D Fermi polarons have also recently been realized in the context of exciton-polaritons in charge-doped atomically thin semiconductors~\cite{Sidler2017}. This scenario has been analyzed~\cite{Sidler2017} using the same type of variational states originally introduced in cold atoms~\cite{Chevy2006}. However, it is currently an open and interesting question whether there is the equivalent of a polaron-molecule transition in the limit of low doping, and this can potentially be answered using the formalism developed in this work.

\acknowledgments 
We are grateful to Yoav Sagi and Zhe-Yu Shi for useful discussions, and to Yoav Sagi for sharing the data of Ref.~\cite{Ness2020}.  JL, WEL and MMP acknowledge support from the Australian Research Council Centre of Excellence in Future Low-Energy Electronics Technologies (CE170100039).  JL and MMP are also supported through Australian Research Council Future Fellowships FT160100244 and FT200100619, respectively.

\appendix

\section{Raman spectroscopy}
\label{app:Raman}
The two-photon transition employed in Raman spectroscopy can be described via the operator
\begin{align}
    \hat{V}_R = \frac{\Omega_R}{2} \sum_\p \left(e^{-i\omega t} \hat{d}^\dag_{\p-\q} \hat{c}_{\p \down} + e^{i\omega t} \hat{c}^\dag_{\p \down} \hat{d}_{\p-\q}  \right), 
\end{align}
where $\hat{d}^\dag_\p$ creates a fermion with momentum $\p$ in the auxiliary impurity state which does not interact with the majority $\up$ fermions. $\Omega_R$ is the Rabi coupling due to the optical field, $\omega$ is the frequency with respect to the bare transition between impurity states, and $\q$ is the imparted momentum. Note that in standard rf spectroscopy, the momentum $\q$ would be essentially zero.

In the case of injection spectroscopy, we start with non-interacting impurities in the auxiliary state and then apply $\hat{V}_R$ to transfer a small fraction of impurities into the $\down$ state. Assuming that the system is initially in thermal equilibrium,  we can use Fermi's golden rule to obtain the transfer rate in the limit of a single impurity:
\begin{align}
    {\cal I}^{\rm R}_{\rm inj}(\q,\omega) 
    = \frac{\pi \Omega_R^2}{2} \sum_\p n_{\rm B}(\p-\q) A_{\rm inj}^{\rm R}(\p;\q,\omega) ,
\end{align}
where
\begin{align} \nn
 & \!\!A_{\rm inj}^{\rm R}(\p;\q,\omega) = \\ \label{eq:spec-inj}
 &  \sum_{n,\nu} \frac{e^{-\beta E_n}}{Z_{\rm med}} |\!\bra{n}\hat{c}_{\p\down}\ket{\nu}\!|^2
 \delta(E_\nu - E_n -\epsilon_{\p-\q}-\omega) .
\end{align}
Here, the non-interacting initial states are taken to be  $\hat{d}^\dag_{\p-\q} \ket{n}$, where $\ket{n}$ are the eigenstates of the $\up$ Fermi gas 
with corresponding energies $E_n$ and partition function $Z_{\rm med} = \sum_n e^{-\beta E_n}$.
For the total interacting system composed of both the $\down$ impurity and the $\up$ Fermi gas,
the eigenstates and associated energies are $\ket{\nu}$ and $E_\nu$, respectively.
Note how we have defined $A_{\rm inj}^{\rm R}(\p;\q,\omega)$ such that the Raman momentum $\q$ only appears in the energy $\epsilon_{\p-\q}$ of the non-interacting impurity state  rather than in the final interacting states. 
Thus, we can directly relate Eq.~\eqref{eq:spec-inj} to the impurity spectral function defined as~\cite{FetterBook}
\begin{align}
    A(\p,\omega) = \sum_{n,\nu} \frac{e^{-\beta E_n}}{Z_{\rm med}} |\!\bra{n}\hat{c}_{\p\down}\ket{\nu}\!|^2
 \delta(E_\nu - E_n -\omega), 
\end{align}
which leads to Eq.~\eqref{eq:spec-rel}. 

Integrating Eq.~\eqref{eq:spec-inj} over $\omega$ 
and removing a complete set of states, we obtain the sum rule in Eq.~\eqref{eq:spec-sum1}
\begin{align}
    \int d\omega \,A_{\rm inj}^{\rm R}(\p;\q,\omega) = \sum_n \frac{e^{-\beta E_n}}{Z_{\rm med}} \expval{\hat{c}_{\p\down} \hat{c}^\dag_{\p\down}}{n} = 1 .
\end{align}
Thus we have
\begin{align}
    \int d\omega \, {\cal I}_{\rm inj}^{\rm R}(\q,\omega) = \frac{\pi \Omega_R^2}{2} \sum_\p n_{\rm B}(\p-\q) = \frac{\pi \Omega_R^2}{2}.
\end{align}
In the main text, 
we define the total spectral function $I_{\rm inj}^{\rm R}(\q,\omega)$ in Eq.~\eqref{eq:Iinj-q} without the prefactor $\pi\Omega_R^2/2$, so that it is normalized to 1 as in Eq.~\eqref{eq:sum1}.

For the case of ejection spectroscopy, where we initially have $\down$ impurities which are then transferred into the non-interacting state, the transfer rate is:
\begin{align}
    {\cal I}_{\rm ej}(\q,\omega) 
    = \frac{\pi \Omega_R^2}{2} \sum_\p A_{\rm ej}(\p;\q,\omega) \equiv \frac{\pi \Omega_R^2}{2} I_{\rm ej}(\q,\omega),
\end{align}
where
\begin{align} \nn
 & \!\!A_{\rm ej}(\p;\q,\omega) = \\ \label{eq:spec-ej}
 &  \sum_{n,\nu} \frac{e^{-\beta E_\nu}}{Z_{\rm int}} |\!\bra{n}\hat{c}_{\p\down}\ket{\nu}\!|^2
 \delta(E_\nu - E_n -\epsilon_{\p-\q}+\omega) .
\end{align}
Here we have a thermal average over the interacting states $\ket{\nu}$, with partition function $Z_{\rm int} = \sum_\nu e^{-\beta E_\nu}$.
Using the properties of the delta function, we thus find
\begin{align}
    A_{\rm ej}(\p;\q,\omega) = \frac{Z_{\rm med}}{Z_{\rm int}} e^{\beta(\omega-\epsilon_{\p-\q})}A_{\rm inj}(\p;\q,-\omega), 
\end{align}
which yields Eq.~\eqref{eq:det-bal1} once we relate the partition functions to the impurity free energy $\Delta F$ via
\begin{align}
    e^{\beta\Delta F}= \frac{Z_{\rm med}Z_{\rm imp}}{Z_{\rm int}}.
\end{align}
Integrating Eq.~\eqref{eq:spec-ej} furthermore yields
\begin{align}
    \int d\omega \,A_{\rm ej}(\p;\q,\omega) = \underbrace{\sum_{\nu} \frac{e^{-\beta E_\nu}}{Z_{\rm int}}  \expval{\hat{c}^\dag_{\p\down} \hat{c}_{\p\down}}{\nu}}_{n_{\rm int}(\p)} ,
\end{align}
which corresponds to Eq.~\eqref{eq:spec-sum2}.

\bibliography{cold-atoms,other-refs}

\end{document}